\documentclass [dvips]{article}
\usepackage{epsfig}
\usepackage{latexsym}
\usepackage{amsfonts}
\usepackage{amsmath}
\usepackage{acronym}
\usepackage{epsfig}
\usepackage{graphics}
\newcommand{\be}{\begin{equation}}
\newcommand{\ee}{\end{equation}}
\newcommand{\bea}{\begin{eqnarray}}
\newcommand{\eea}{\end{eqnarray}}
\makeatletter
\@addtoreset{equation}{section}
\def\theequation{\thesection.\arabic{equation}}
 \makeatother
\begin{document}
\normalsize
\title{\Large Fermions with spin $1/2$ as global $SO(3)$ vortices.}
\author
{{\bf L.~D.~Lantsman}\\
 Wissenschaftliche Gesellschaft bei
 J$\rm \ddot u$dische Gemeinde  zu Rostock,\\Augusten Strasse, 20,\\
 18055, Rostock, Germany; \\ 
Tel.  049-0381-799-07-24,\\
llantsman@freenet.de}
\maketitle
\begin {abstract}
In this paper we show that the nontrivial fundamental group $\pi_1 SO(3) ={\Bbb Z}_2$ for the  group $SO(3)$ of global proper rotations of a four-dimensional Euclidian space (when a spin structure is introduced preliminarily in that space) implies always fermions as global $SO(3)$ vortices, while bosons can be reduced to trivial lines (contracted into a point) in the $SO(3)$ group space.
\end{abstract}
\newpage
\tableofcontents
\newpage
\section{Introduction.}
There is no special need to discuss here the entire importance for modern physics subdividing quantum fields into two  categories, bosons and fermions. But revealing the source of such subdividing is the very important and interesting task.

In the present study we propose the simple and enough transparent way to understand the distinction between two kinds of particles spins: integer and half-integer. 

The  source of such distinction we see in the nontrivial topological structure of the   group $SO(3)$ of global proper rotationsin of a four-dimensional Euclidian space (the pattern of such spaces is the Minkowski space $M$).

The mentioned nontrivial topological structure of the   group $SO(3)$, the natural subgroup in the general Lorentz group, comes to the nonzero fundamental group $\pi_1 SO(3) ={\Bbb Z}_2$ of one-dimensional loops in its group space.

In this case, as it was discussed in Ref. \cite{Penrous1}, this two-connection of $SO(3)$ implies the existence of two kinds of loops in the group space.
Firstly, there are loops which can be contracted into a point. Such loops lie inside the sphere $S^3\simeq SO(3)$ without intersecting its poles. And the second kind of loops just includes the poles of $S^3$. As a consequence, antipodal points of the sphere $S^3$ can be identified, $x\sim-x$ ($x$ are points of the given Euclidian four-dimensional space),  displaying the natural isomorphism between the $SO(3)$ (global) group space and the projective space ${\bf RP}_2$. 

This means, in particular, that at the  rotation 
onto the angle $2\pi$ around a loop belonging to the "second type", any spinor object $\varphi(x)$ (with the spin $1/2$) changes its sign onto the opposite one \cite{A.I.}, and only the  rotation 
onto the angle $4\pi$ around such a loop returns the spinor object $\varphi(x)$ in its initial place \cite{Penrous1}.

Vice verse, for "first type" loops, it is  sufficient  rotations 
onto the angle $2\pi$ to return spin-vectors $\chi_{s\mu}$ (if particles with spins $1$ are in the question) \cite{A.I.} in their initial places.

\medskip The said shows transparently the presence of two types of loops and, that is the same, of two types of spinors (if the spin structure is specified in the given four-dimensional Euclidian space), associated with the natural  two-connection \cite{Penrous1} of the global $SO(3)$.

From the topological standpoint, the evident impossibility \cite{Penrous1} to deforme "first type" loops to "second type" ones means a domain wall between two topological sectors of the global $SO(3)$:
\be\label{dom}  \pi_0 ~ SO(3)=  \pi_1 ~ SO(3)={\Bbb Z}_2. \ee
On the other hand, the topological chain (\ref{dom}) implies \cite{Al.S.} the existence of ${\Bbb Z}_2$ (global) vortices associated with the global $SO(3)$ group. These global vortices are just fermionic fields.

\section{Why setting spin structure in an Euclidian manifold always implies global vortices?}
\subsection{Isotropic flags and spin-vectors.}
Following \cite{Penrous1}, let us consider an isotropic vector $\bf K$ in the Minkowski space. With O being the origin of coordinates in the Minkowski space, we choose ${\bf K}\equiv \overrightarrow {O~R}$.

On the other hand, directing an isotropic vector $\bf K$ onto the past/future, one subdivides finally the Minkowski space $M$ into two subspeces: respectively ${\cal G}^-$ and ${\cal G}^+$.  These subspeces in a frame $(T,X,Y,Z)$ can be represented by the intersections $S^-/S^+$ of the past/future light cone
\be \label{light cone} 
T^2-X^2-Y^2-Z^2=0
\ee
with the hyperplanes $T=-1$ ($T=1$).

In the flat space $M$ the mentioned intersections are, indeed, the spheres (see \cite{Penrous1} and also \cite{rem2})
\be \label{light cone1}
x^2+y^2+z^2=1
\ee 

\newpage
 \bigskip
\begin{figure}[h]
\centering			
		\includegraphics[bb= 250 250 300 300]{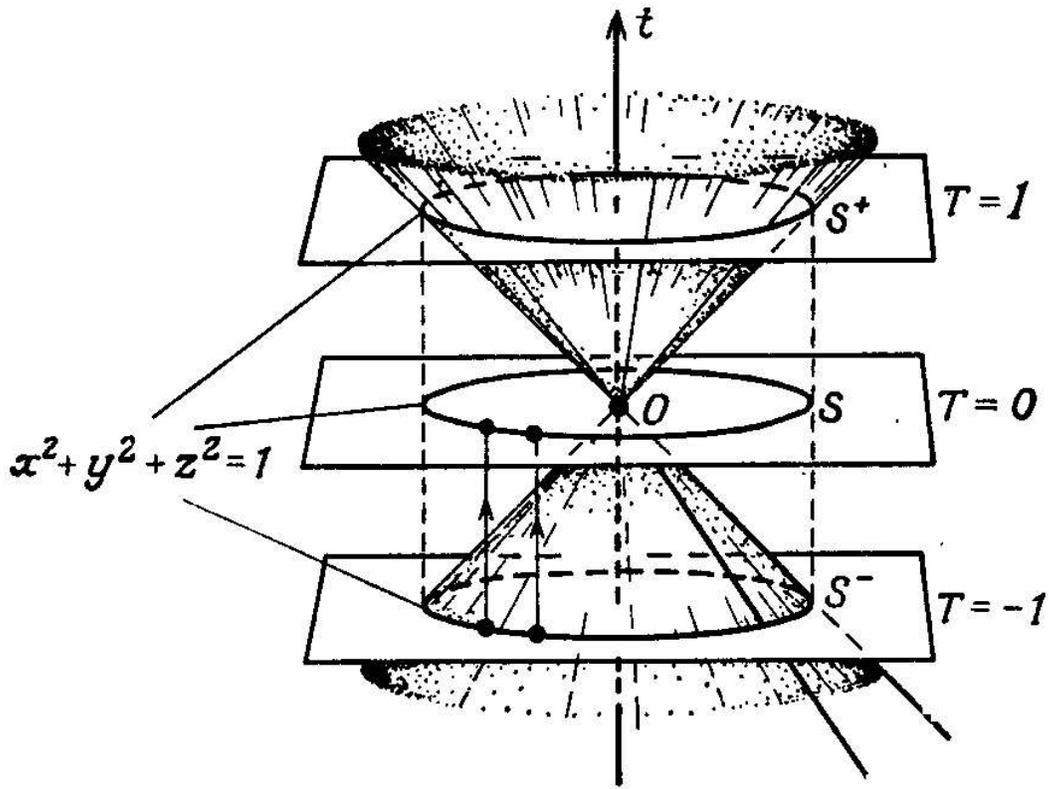}\hspace{20.mm} \vspace{-100.mm}
\caption{The Minkowski space $M$ can be subdivided into celestial, $S^-$, and anticelestial, $S^+$, spheres with respect to the observer.}				
	\label{fig:p24}
\end{figure}
\newpage
\bigskip
\par We see from Fig. 1 above that the internal part of the sphere $S^-$ represents the set of time-like directions of the past, while the internal part of the sphere $S^+$ represents the set of time-like directions of the future. The parts of the hyperplanes $T=-1$ ($T=1$) out the mentioned spheres represent space-like directions.

Let now an observer be located in the origin of coordinates $O$. Light beams passing through his eye correspond in this case to isotropic straight lines  passing through $O$, while the past directions of the mentioned lines form the field of vision of the observer.

It is just the space ${\cal G}^-$, can be represented correctly by the sphere $S^-$. Actually, $S^-$ is the exact geometrical image of that the observer can "see" at the condition he is immovable with respect to the reference frame $(t,x,y,z)$, i.e. that his world velocity is $ct$.

Indeed, the observer can think he is permanently in the centre of a unit sphere $S$ (his sphere of vision) on which he maps all he sees in any time instant. The straight lines going from his eyes to these points of $S$ are the projection of worlds lines of coming beams on his "instanton" space $T=0$.

Thus the mentioned images are congruent to the images in $S^-$ (see Fig. 1). The reasoning  just performed allows to refer to the space ${\cal G}^-$ ($S^-$) as to the {\it celestial} sphere of the point $O$ \cite{Penrous1}. The map of past isotropic directions let out from $O$ in the points of $S^-$ was called {\it the celestial map} in Ref. \cite{Penrous1}.

Since any isotropic vector $\bf L$ directed in the past is connected, in the unique and relativistic invariant wise, with an isotropic vector directed in the future (it is the vector $-{\bf L}$), the field of vision of the observer can be represented also by the sphere $S^+$. Such representation can be called the {\it anticelestial map} \cite{Penrous1}.

The correspondence between $S^+$ and $S^-$ it is merely the correspondence $(x,y,z)\to(-x,-y,-z)$, i.e. it is the diametrally opposite correspondence at the superposition of the  one sphere onto another. Such a map changes the orientation of the sphere onto the opposite one.

\bigskip The sphere $S^-$ ($S^+$) can be considered, in a natural way, as a {\it Rimanian sphere} \cite{Penrous1} of the Argand plane (or of the {\it Argand-Bessel-Gauss plane}); this sphere is the well known representation for complex numbers including infinity. The ordinary properties of the Argand plane and its Rimanian sphere reflect various geometrical properties of the Minkowski space $M$. In particular any restricted Lorentz transformation \footnote{It is the Lorentz transformation maintaining the spatial and the time orientations of the Minkowski space $M$ \cite{Penrous1}.} proves to be specified uniquelly by its action onto the Rimanian sphere (and thus onto isotropic directions).

One can replace the coordinates in the sphere $S^-$ with a one complex number got by means of the "steriographical" correspondence between the sphere $S^-$ and the Argand plane (see Fig. 2).

Let us consider the  plane $\Sigma$ set by Eq. $z=0$ in the Euclidian 3-space $T=1$ and let us map the points of $S^+$ onto this plane $\Sigma$ by means of projecting the nord pole $N(1,0,0,1)$. Let $P(1,x,y,z)$ and $P'(1,X',Y',Z')$ are proper points on $S^+$ and $\Sigma$. Denote then as $A$ and $B$ the finite points of the perpendiculars droped from $P$ into $CP'$ and $CN$. Labeling the  points in $\Sigma$ by the complex parameter
\be \label{ksjusha}
\zeta=X'+iY',
\ee
we get
\be \label{podobie}
x+iy=h\zeta,
\ee
where
\be \label{podobie1}
h=\frac {CA}{CP'}=\frac {NP}{NP'}
=\frac {NB}{NC}=1-z,
\ee
that follows from the fact similarity of the triangles $NBP$ and $NCP'$.

Whence the  parameter $\zeta$ can be expressed as
\be \label{ksjusha1}
\zeta=\frac {x+iy}{1-z}
\ee
through the coordinates $(1,x,y,z)$ of the point $P$.

To get the inverse relation, we should exclude $x$ and $y$ from (\ref{ksjusha1}) taking account of (\ref{light cone1}):
\be \label{xixi}
\zeta \bar \zeta =\frac{x^2+y^2}{(1-z)^2}=\frac{1+z}{1-z}.
\ee
Solving (\ref{xixi}) respectively to  $z$ and substituting the expression has been got in (\ref{ksjusha1}), we have
\be \label{xyz}
x=\frac{\zeta +\bar \zeta}{\zeta \bar \zeta+1}, \quad y=\frac{\zeta -\bar \zeta}{i(\zeta \bar \zeta+1)}, \quad z=\frac{\zeta \bar \zeta-1}{\zeta \bar \zeta+1}.
\ee
The relations (\ref{ksjusha1}) and (\ref{xyz}) set the standard steriographical correspondence between the Argand plane $\zeta$ and  the unit sphere in the $(x,y,z)$-space with its centre in the point $(0,0,0,0)$. It is the one-to-one correspondence if one think that $\zeta=\infty$ is the one "point" added to the Argand plane and herewith associated with the nord pole of the sphere.

Thus the sphere $S^+$ gives the standard realization of the Argand plane $\zeta$ with the added point $\zeta=\infty$; it represents correctly the {\it Rimanian sphere} \cite{Penrous1} $\zeta$.
\bigskip

\begin{figure}[h]  
	\label{fig:p25}
	\centering
	\includegraphics[bb= 200 200  300  300]{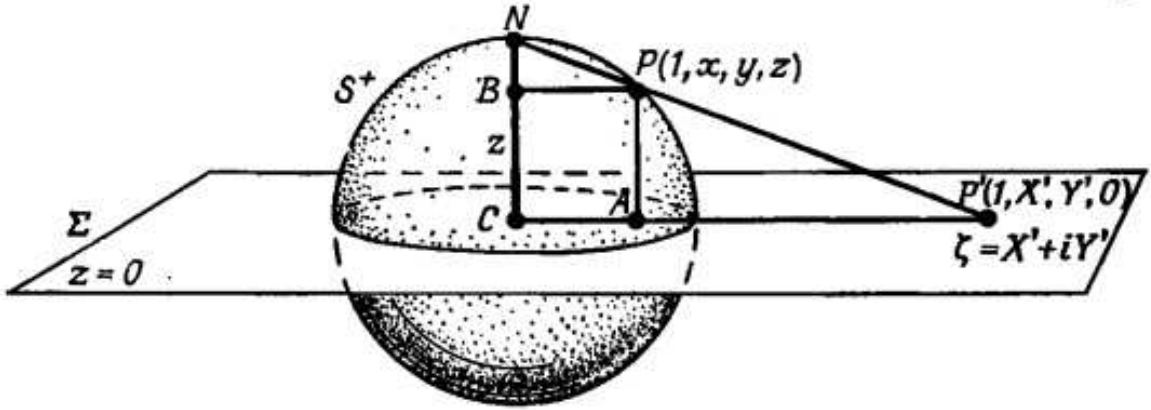}
	\vspace{50.mm} \hspace{60.mm}
\caption{The steriographical projection of the $S^2$ sphere into the Argand plane.}	
\end{figure}
\bigskip
To avoid the infinite coordinate $\zeta=\infty$ for the point $(1,0,0,1)$ in the nord pole of the sphere $S^+$, it is convinient sometime to label the points on $S^+$ with the pair $(\xi,\eta)$ of complex numbers (not equal to zero simultaneously) obeying the condition \cite{Penrous1}
\be \label{zet}
\zeta=\xi/\eta.
\ee
Such complex numbers are the {\it projective} (homogenious) complex coordinates \cite{Al.S.}; thus at an arbitrary different from zero complex number $\lambda$, the pairs $(\xi,\eta)$ and $(\lambda\xi,\lambda\eta)$ image the same point on $S^+$. In these coordinates the additional point on the infinity, $\zeta=\infty$, is set by the finite label, for instance $(1,0)$.
Thus one can consider $S^+$ as a realization of a {\it complex straight projective line} ${\bf CP}_1\simeq {\bf RP}_2$.

In these complex homogenious complex coordinates Eq. (\ref{xyz}) acquires the look \cite{Penrous1}
\be \label{xyz1}
x=\frac{\xi\bar\eta+\eta\bar\xi}{\xi\bar\xi +\eta\bar\eta},\quad y=\frac{\xi\bar\eta-\eta\bar\xi}{i(\xi\bar\xi +\eta\bar\eta)}, \quad z=\frac{\xi\bar\xi -\eta\bar\eta}{\xi\bar\xi +\eta\bar\eta}.
\ee
Remember now that the role of the point $P(1,x,y,z)\in S^+$ comes merely to representing an isotropical direction going from the origin $O$. Indeed, one can choose another point on the straight  line $OP$; this point also represents correctly the same isotropical direction.  For instance, it can be a point $R\in OP$ with the coordinates $(T,X,Y,Z)$ got from the coordinates of $P$ by multiplying on
$$ \frac{ \xi\bar\xi+ \eta\bar\eta}{\sqrt 2}, $$
where the multiplier $1/\sqrt 2$ was introduced for the convenience.

Now the vector ${\bf K}:= \overrightarrow {O~R}$ has the coordinates
$$
T=\frac 1{\sqrt 2}(\xi\bar\xi+ \eta\bar\eta), \quad   X=\frac 1{\sqrt 2} (\xi\bar\eta+\eta\bar\xi),
$$
\be \label{bfK}
Y=\frac 1{i\sqrt 2} (\xi\bar\eta-\eta\bar\xi), \quad Z=\frac 1{\sqrt 2}(\xi\bar\xi -\eta\bar\eta).
\ee
\medskip
Thus the complex pair $(\xi,\eta)$ can be always associated with an isotropic vector ${\bf K}:= \overrightarrow {O~R}$ directed in the future. On the other hand, these complex coordinates are redundand for $\bf K$ since a phase transformation $\xi\to e^{i\theta}\xi$, $\eta\to e^{i\theta}\eta$ retains $\bf K$ immovable.

Now we desire (following \cite{Penrous1}) to connect with $(\xi,\eta)$ a richer geometrical structure at which this redundance comes to the unique (but essential) uncertainty in the sign. In turn, such a structure is reduced actually \cite{Penrous1} to the {\it isotropic flag} , i.e. to the  isotropic vector ${\bf K}$, representing $\xi$ and $\eta$ to within a phase, and to the {\it cloth of the flag}, i.e. to the isotropic  half-plane attached to $\bf K$ and representing the phase.

If the phase angle changes onto $\theta$, the flag is turned onto $2\theta$, that implies the mentioned uncertainty in the sign.

The important claim to any geometrical image of the complex pair $(\xi,\eta)$ consists in its independence on the utilized coordinates. If a pair $(\tilde\xi,\tilde\eta)$ is got from $(\xi,\eta)$ by means of a spin transformation 
\cite{Penrous1}
\be \label{spin transformation}\left(
\begin{array}{llcl}
T+Z\quad X+iY\\ X-iY \quad T-Z
\end	{array}
\right ) \longmapsto\left(\begin{array}{llcl}
\tilde T+\tilde Z\quad \tilde X+i\tilde Y\\ \tilde X-i\tilde Y \quad \tilde T-\tilde Z
\end	{array}\right ) = {\bf A} \left(
\begin{array}{llcl}
T+Z\quad X+iY\\ X-iY \quad T-Z
\end	{array}
\right )
{\bf A}^*
\ee
where ${\bf A}$ is the unimodular matrix
\be \label{unimod}
{\bf A}:=\left(
\begin{array}{llcl} \alpha\quad\beta \\ \gamma   \quad \delta \end	{array}\right ); \quad {\rm det}~{\bf A}=1,
\ee
and ${\bf A}^*$ is the matrix complex conjugate and transposed to ${\bf A}$, then an abstract spin-vector $\kappa$ represented by the pair $(\xi,\eta)$ remains immovable due to (\ref{bfK}).

Thus if a pair $(\xi,\eta)$ sets a geometrical representation of a spin-vector $\kappa$ in a one coordinate system in the Minkowski space, then the pair $(\tilde\xi,\tilde\eta)$ would set the same structure in the second, transformed, coordinate system \footnote{It is just the {\it passive} Lorentz transformation \cite{Penrous1} $G:U^i\mapsto U^{\hat i}$ ($i=0,1,2,3$) for a  tetrad $U^i$ in the Minkowski space ($U^i$ are the coordinates of the vector $\bf U$ in this space) we shall return to this topic in the next subsection.}.

\bigskip To understand how to go over from a flag to the apropriate spin-vector, one would to clarify the nature of the uncertainty in the sign for the representation of the isotropic flag by the pair $(\xi,\eta)$.

For this purpose, let us consider the action of the transformations 
\be \label{lambda}
(\xi,\eta)\longmapsto (\lambda\xi,\lambda\eta)
\ee
onto an isotropic flag (here $\lambda \neq 0$ is a complex number). Such transformations maintain the direction of the  flagstaff, but they can change its  extent or the direction of the cloth of the flag.

Let us set
\be \label{lambda1}
\lambda=r e^{i\theta},
\ee
where $r,\theta\in {\bf R}$ and $r>0$. Then if $\theta=0$ (i.e. at real $\lambda$) the transformation (\ref{lambda}) maintains invariant the cloth of the flag, while the extent of the flagstaff increases acquiring the multiplier $r^2$ (this can be checked directly at substituting (\ref{lambda}) in (\ref {bfK})).

Simultaneously, if $r=1$ (i.e. if $\vert \lambda \vert=1$), the transformation (\ref{lambda}) das not affects the flagstaff but the cloth of the flag is turned onto the angle $2\theta$ in the positive direction.

It can be explained simpler at considering two  infinitely close points $P$ and $P'$ on $S^+$. Let $P$ is given by the coordinate $\zeta$ while $P'$ by the coordinate $\zeta-2^{-1/2}\epsilon \eta^{-2}$. As a result of the transformation (\ref{lambda}), we have $\eta\to\lambda\eta$, whence 
$$\eta^{-2}\longmapsto r^{-2} e^{-2i\theta} \eta^{-2}.$$
Since the extention of the flagstaff ischanged inversely proportionally to the infinitelysimal separation $PP'$, the first part of our assertion is proved.

The second part of our assertion follows from the above discussed fact that the sphere $S^+$ is got from the Argand plane $\zeta$ as a result of the conformal stereographical projection.

Let us consider the continuous rotation
$$ (\xi,\eta) \longmapsto  (e^{i\theta}\xi,e^{i\theta}\eta), \quad \theta \in [0,\pi].$$
We get finally
\be \label{minus}
(\xi,\eta) \longmapsto (-\xi,-\eta),
\ee
but the flag returns to its initial position; herewith the cloth of the flag turns onto the angle $2\pi$ (i.e. it makes the complete revolution around the flagstaff).

Continuing the  rotation in such a wise that $\theta$ will vary in the interval $[\pi,2\pi]$, we get once again the initial pair $(\xi,\eta)$. Thus to return $(\xi,\eta)$ to its initial position, it is necessary to turn the cloth of the flag onto the angle $4\pi$.

This reasoning shows that the complete local geometrical representation of the pair $(\xi,\eta)$ in the Minkowski space with account of its sign is {\it impossible}. Any local structure in the Minkowski space which one attempt to asociate with an isotropic flag also will turn onto the angle $2\pi$ and thus return to its initial position at the transformation (\ref{minus}).

To see this more clear, note firstly that one can perform a change
\be \label{spt} (\xi,\eta) \longmapsto  (e^{i\theta}\xi,e^{i\theta}\eta)\ee
by means of a spin transformation corresponding to a rotation at which the direction of the flagstaff is the invariant isotropic direction \footnote{For simplicity, it can be chosen
$$ (\xi,\eta)=(0,1) \longmapsto (0,e^{i\theta}) $$
}.

Since $\theta$ varies continuously from 0 to $\pi$, a spin transformation \footnote{A general (nonsingular, unimodular) complex linear  spin transformation of the coordinates $\xi$ and $\eta$ has the look \cite{Penrous1}
$$ \xi  \longmapsto \tilde \xi=\alpha\xi+\beta\eta,$$
$$  \eta  \longmapsto \tilde  \eta =\gamma \xi+\delta\eta.$$
In the matrix shape this (unimodular) transformation can be rewritten as
$$ \left (
\begin{array}{llcl} \tilde \xi\\ \tilde  \eta\end{array}\right )={\bf A} \left (
\begin{array}{llcl}  \xi\\   \eta\end{array}\right ).
$$} is also changed continuously (at the condition that the rotation axis is fixed), and finally it comes to the transformation $-{\bf I}$.

The appropriate Lorentz transformation is also changed continuously, but it is finished by the identical Lorentz transformation. Thus {\it any} geometrical structure on the Minkowski space $M$ would return to its initial position course its rotations, in spite the pair $(\xi,\eta)$ is transformed into  the pair $(-\xi,-\eta)$ course these rotations.

Since, as it was just established, any complete local geometrical representation in $M$ is impossible, it becomes obvious how we shall act now. We should expand the notion of the geometry in the Minkowski space $M$ toward the "legalization" of those values which don't return to their initial positions at the rotation around an axis on the angle $2\pi$, but these values {\it would} return to their initial positions at the rotation around this axis on the angle $4\pi$. Such values are called {\it the spinor objects} \cite{Penrous1}. 

In particular, a spin-vector differs from an isotropic flag only as a spinor object. {\it Two and only two} spin-vectors correspond to this isotropic flag.
\subsection{Geometrical specifying  spin-vectors.}
Now we are abble to give the geometrical definition of a spin-vector. We shall think that $Q$ is an isotropic flags on the Minkowski space $M$ while $\cal E$ is the whole space of isotropic flags. 

We should make sure that the space $\cal E$ possesses indeed the necessary topological properties. Since it is four-dimensional, it cannot be topologically equivalent to the space $SO(3)$ (the latter one is the three-dimensional space) or to $O_+^\uparrow (1,3)$ (the latter one is the six-dimensional space) \footnote{This  will be discussed in Appendix 1.}. 

 Nevertheless, as in the $O_+^\uparrow (1,3)$ case, the {\it essential} part of the topology of the considered space is the same as in the $SO(3)$ group space \footnote{One has ${\cal E}\cong SO(3)\otimes {\Bbb R}$; threfore $\pi_1{\cal E}={\Bbb Z}_2$.}. 
 
 To make sure in the said, one can consider the $S^+$-representation. Any element $Q$ of the space $\cal E$ can be represented by a point $P$ on  $S^+$ an a nonzero tangential vector $\bf L$ to $S^+$ in $P$. In a continuous (but not an invariant) wise, one can associate with $Q$ a Cartesian reference frame  by choosing  the axis $z$ to be directed from the origin of coordinates in the point $P$,  the axis $x$ to be parallel to  $\bf L$ and the axis $y$  to suplement this reference frame.  
 
 Such a  reference frame corresponds unambiguous to  points of the space $SO(3)$.  The only free parameter characterizing $Q$ is $\| L \|$, and  this parameter is indeed a positive real number, being simultaneously topologically  invariant. Whence $\cal E$ possesses the requested properties. 
 
 \medskip We assume that the space $\cal E$ possesses the {\it two-fold} universal covering $\tilde {\cal E}$.  We claim herewith  that two different images $Q_1, Q_2\in \tilde {\cal E}$ of a $Q\in {\cal E}$ changes by their places at the continuous rotation onto the angle $2\pi$. 
 
 More exactly, any isotropic flag $Q$  sets two spin-vectors $\kappa$ and $-\kappa$ in $\tilde {\cal E}$. Any  continuous rotation onto the angle $2\pi$ will transfer $\kappa$ in $-\kappa$, and since  $-\kappa$ returns back  into $\kappa$, we write 
 \be\label{kappa} 
 -(-\kappa)=\kappa.
 \ee
 In addition,  there exists the unique {\it zero spin-vector}, denoting as ${\bf 0}$ \cite{Penrous1}, which does not  correspond to any flag.  The zero spin-vector is associated with the zero world vector playing the role of the ``flagstaff'' while the ``flag cloth'' is not specified. 
 
 A pair $(\xi,\eta)$ can be treated indeed as the components of the spin-vector $\kappa$. The  spin transformations applied to the  pair $(\xi,\eta)$ will correspond to the active motions,  transforming $\kappa$ relatively the Minkowski space $M$. 
 
 A continuous rotation onto the angle $2\pi$ corresponds to the  sequence of spin transformations acting on $(\xi,\eta)$ and leading to $(-\xi,-\eta)$. Thus the pair $(-\xi,-\eta)$ represents in fact the  components of the spin-vector $-\kappa$. 
\subsection{Topological specific of $SO(3)$ group space.}
As it is well known, "proper" spatial rotations in the Minkowski space $M$ form the group $SO(3)$ consisting of $3\times 3$ orthogonal matrices with unit determinants.

The $SO(3)\simeq S^2$ group manifold can be utilized for the representation of different orientations of a geometrical object in the Minkowski space $M$. If one chooses any such orientation as the initial orientation, representing it by the unit element of $SO(3)$, another element of $SO(3)$ will represent the  orientation got from the initial one by means of the appropriate proper spatial rotation.

Any such rotation is determined by its rotation axis $\bf k$ and the right-handed rotation on the angle $\theta$. Therefore it can be represented by the vector of the length $\theta$ in the direction $\bf k$. Since we can consider only the interval $\theta\in[0,\pi]$, any point of the $SO(3)$  group manifold corresponds to the point of the closed ball $B$ with the radius $\pi$.

However this correspondence is not a one to one since a rotation on the angle $\pi$ with respect to the rotation axis $\bf k$ represents the same that the rotation on the angle $\pi$ with respect to the rotation axis $-{\bf k}$. Identifying the opposite points of the boundary $S^2$ of the ball $B$, we get the space $\hat B$ representing rotations in the unique and continuous way (in other words, intimate points of the space $\hat B$ represent rotations differing insignificantly from each other).

\medskip Our interest now is the topology and especially the question about the connection in the space $\hat B$. 

As it is well known, a space is called {\it the one-connected} if any closed contour therein can be contracted into a point. In particular, it is correctly, obviously, for an Euclidian space, for a spherical surface (concretely, for $S^k$ with $k\geq 2$ \cite{Al.S.}), for an Euclidian space with a removed point.

Mathematically, the "one-connected" nature of a space can be expressed as following \cite{Penrous1}: if $c_1$ and $c_2$ are two open curves joining two points of this space, then the curve $c_1$ can be deformed continuously into the curve $c_2$.

The space $\hat B$ {\it is not} one-connected. Closed contours in this space are subdivided into two different classes I and II depending on that they have either odd or even number of "intersections" with $S^2$. Any such intersection occurs if a curve reaches $S^2$ and is repeated on the diametrically opposite end (one can make sure in this by identifying the points).

All the diameters of the ball $\hat B$ belong to the class I. All the internal contour belong to the class II; in particular, there are "trivial" contours consisting of a one point. Herewith none contour of the class I can be deformed continuously into a contour of the class II since the intersection points with $S^2$ can arise and disappear only in pairs. On the other hand, {\it all the} contours of the class I can be deformed continuously into each other; the analogous assertion is correct  also for contours of the class II. The cause of the latter assertion  again the possibility to exclude in pairs the intersection points with $S^2$ (in Fig.3 it is depicted the method how to do this step by step), while all the internal contours just as those intersecting $S^2$ once can be deformed one in another.

\bigskip
\begin{figure}[h]  
	\label{fig:p26}
	\centering
	\includegraphics[bb=   249  249 230 230]{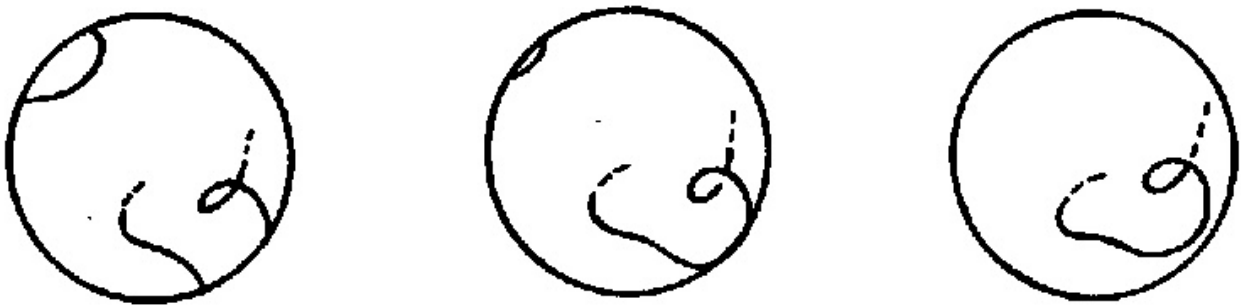}\hspace{120.pt}\vspace{210.pt}
\caption{The $SO(3)$ group space is the closed 3-ball the diametrically opposite points of which are identified. by means of a continuous deformation of a curve in $SO(3)$, the pairs of intersections with $S^2$ can be removed}	
\end{figure}
\medskip
Now let us consider a continuous rotation of an object in  the Euclidian 3-space bringing this object to its initial orientation. Such a rotation corresponds to the closed contour in  the $SO(3)$ group manifold (and therefore also in  the space $\hat B$) referring either to the class I or to the class II.

It is obvious that in  the case of a simple rotation on the angle $2\pi$ one gets the  contour of the class I, while a rotation on the angle $4\pi$ results a contour of the class II. It becomes evident from the said above that a rotation on the angle $2\pi$ (where the {\it complete} motion would be taken into account and not only its initial and final orientations) cannot be deformed continuously into a trivial motion corresponds to absence of any rotation. Simultaneously, any rotation on the angle $4\pi$ {\it can be} reduced to a trivial motion.

The important point in  the above reasoning is considering implicating the ball $\hat B$.

\bigskip
There are lot of ways to  illustrate the result just obtained.
 
The one of ways to perform a continuous deformation between a  rotation on the angle $4\pi$ and the "trivial rotation" (i.e. to absence of any rotation) is the following (H. Weyl). Let us consider a pair of  right cones with the equal semiangles $\alpha$ in the Euclidian 3-space, herewith the one of  these cones is fixed while the second  rolls freely along the fixed one in such a wise that their apices remain combined. We begin  from an infinitesimal $\alpha$ and let us then drive the mobile cone the one time around the fixed one in such a wise that the mobile cone turns onto the angle $4\pi$.

Let $\alpha$ to increase gradually from 0 to $\pi/2$. At each fixed $\alpha$, one observes a circular motion since the mobile cone turns one time around the fixed one. But when $\alpha$ approaches $\pi/2$, the considered cones become almost flat while the motion turns into a simple contact of these cones. Thus at $\alpha=\pi/2$ we get a ``trivial'' contour in the $SO(3)$ group space and rotations on the angle $4\pi$ can be deformed continuously into  a trivial rotation, corresponding to the rest state.

\medskip
In the well-known {\it Dirac   puzzle} with the scissors a lace is  put through a one ring of the scissors,  then it is passed behind a one  post of the   chair's back,  put through the other ring passed behind the other chair's back;  then its ends are bound together.

\newpage
\begin{figure}[h]  
	\label{fig:p27}
	\centering
	\includegraphics[bb= 170 170  130  130]{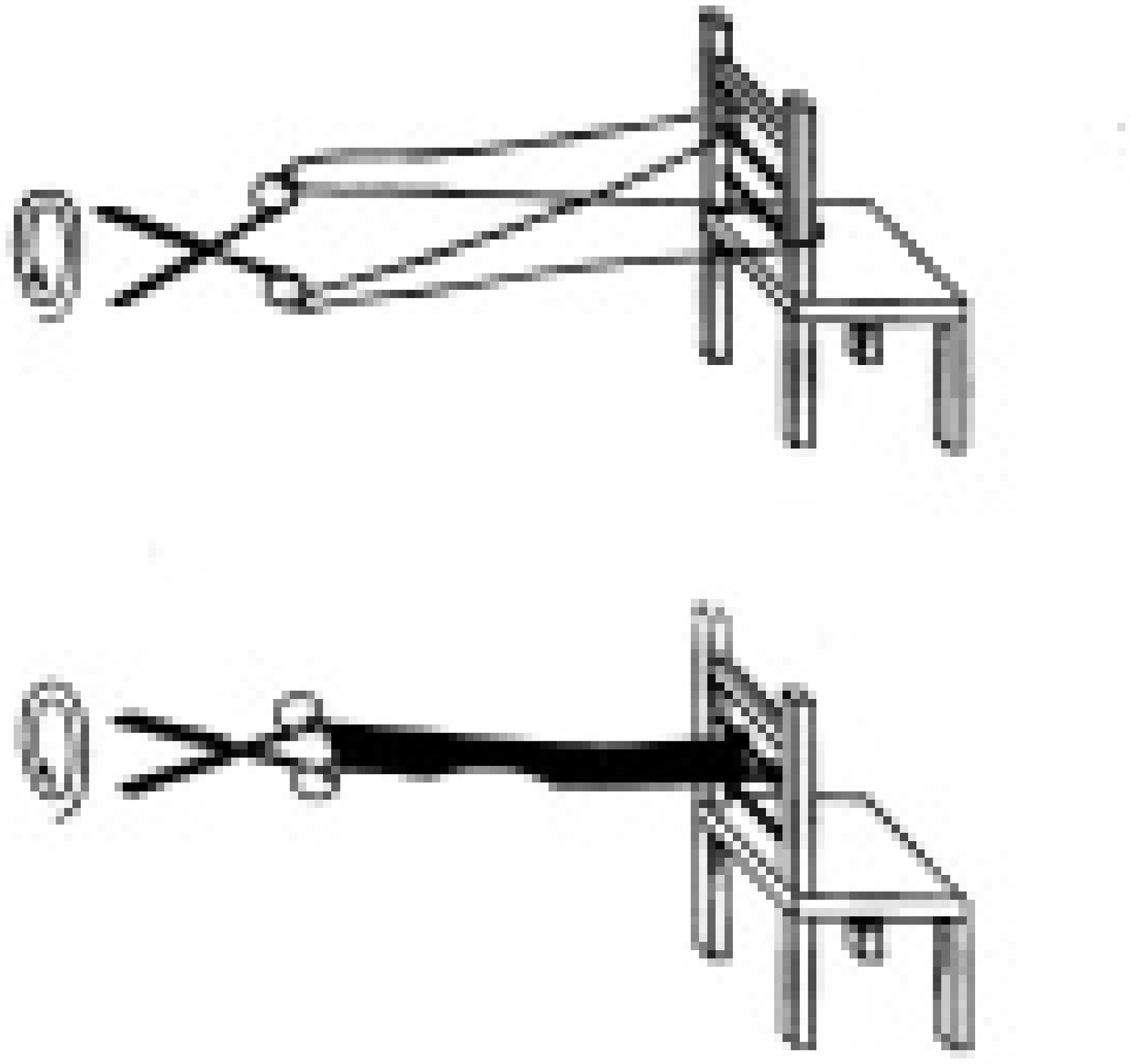}\vspace {-400. pt}\hspace{400.pt}
\caption{The Dirac   puzzle with the scissors. Turn the scissors on $720^\circ$; after this untangle the lace without moving the chair and without revolving  the scissors. With the tape, it can be performed much simpler.  }	
\end{figure}
\newpage 
Let us turn the scissors on the angle $4\pi$ with respect to its symmetry axis and let us  propose to someone to untangle the lace without revolving  the scissors and without moving the chair. The circumstance that this task can be solved for the angle $4\pi$ but not for $2\pi$ \footnote{Indeed, the proof of the fact that it cannot be solved for the angle $2\pi$ requieres a profound topological analysis \cite {Newman}.} is the consequence of the above discussed properties of the $SO(3)$ group manifold.

The solution  becomes trivially simple if four segments of the lace (the latter one is  necessary only to complicate the matter) are represented as those  sticked (in an arbitrary wise) to a tape hooked  on the chair: roling the tape onto the angle $4\pi$ will be untangled if one will  encircle by the middle part around its free end. 


\bigskip The connection of the space $\hat B$ can be investigated by means of considering  ``open'' curves linking the points $P$ and $Q$. Again (for the fixed $P$ and $Q$) these curves are subdivided into two classes, I and II, with respect to even or  odd number of their intersections with $S^2$. And again any  curve belonging to the fixed class can be deformed continuously into another curve of this  class, but it  cannot be deformed continuously into a curve belonging to another class. 

The proof of this statement is analogous to the above one but with  the distinction that there are no essential topological differentiation between the classes I and II (in the case of {\it closed} contours, the differentiation between the classes I and II is not an essential topological differentiation: all the  contours of the  class II and only of this  class can be contracted into a point.)

From the $\hat B$ topology standpoint, such a situation  appears because the concrete   position of the boundary  $S^2$ is not important: for example, one can  think that the ball $B$ is located outside of $B$, and  then we  move the boundary $S^2$ of the ball $B$  in the one radial direction  outside the ball $B$ and in the opposite direction, inward the ball $B$. If a  curve linking the points $P$ and $Q$ intersects $S^2$ one time in the initial position, it, generally, will not intersect $S^2$ in the final position. 

Note also that two curves linking the points $P$ and $Q$ belong to the one fixed class if and only  if the first  together with the second one, following after the former in the opposite direction, form a closed contour of the class II (i.e. that can be contracted into a point). 

Returning to the initial Euclidian three-space, it is worth to remark that the points $P$ and $Q$ correspond to two orientations $\cal R$ and $\cal Q$ of the one object while the way from $P$ in $Q$ in the space $\hat B$ corresponds to the continuous motion begining with the orientation $\cal R$ and ending with the orientation $\cal Q$. 

However,  there are two in essential different classes of continuous motions between $\cal R$ and $\cal Q$. The motions belonging to the definite class  can be deformed continuously into each other but they cannot be deformed in any motion belonging to the second class. Nevertheless, there are no internal property allowing to distinguish between the above classes. 

\bigskip The topological specific of the $SO(3)$ group space, discussed in the present subsection, is connected with its {\it fundamental group}
\be \label{funs} 
\pi_1 SO(3)={\Bbb Z}_2.
\ee
It will be useful and  cognitive to consider here another examples of manifolds $M$ implicating the fundamental group $\pi_1 M={\Bbb Z}_2$.

The typical such case is the case \cite{Al.S.} of liquid nematic crystals possessing a one symmetry axis directed along the axis $z$ in the chosen (Cartesian) coordinate system. \par
In this case the initial
$$ SO(3) \simeq SU(2)/\mathbb{Z}_2$$ 
(rigid) symmetry of such a liquid nematic crystal is violated thereupon 
down to its $O(2)$ subgroup; thus the appropriate degeneration space proves to be 
\be 
\label{nem} R_n=  SO(3)/O(2)\simeq S^2/{\bf Z}_2 \simeq {\bf RP}^2
\ee
since 
$$ SO(3)/SO(2)\simeq S^2$$ 
and
$$ O(2) \simeq SO(2)\otimes {\Bbb Z}_2.$$ 
From the thermodynamic standpoint \cite{Al.S.}, upon violating the initial $SO(3)$ symmetry in a liquid nematic crystal possessing a one symmetry axis,  the free energy $F$ of this crystal attains its minimum just over the degeneration space $R_n$. 

On the other hand, the degeneration space $R_n$ contains the specific type of topological defects, {\it the disclinations} \cite{Al.S.}.  The cause of disclinations is in the isomorphism \cite{Al.S.}
\be \label{proect1}\pi_0 (O(2))= \pi_1 R_n=\pi_1 ({\bf RP}^2)={\Bbb Z}_2\ee 
(the group $O(2)$ consists of orthogonal $2\times 2$ matrices with determinants $\pm 1$; this just implies that the $O(2)$ group space is two-connected).

\medskip Note   that the topological equality 
\be\label{funp} \pi_1 ({\bf RP}^2)={\Bbb Z}_2\ee 
is equivalent to the topological equality (\ref{funs}) and thus can be explained by arguments similar to those   \cite{Penrous1} we  have utilized in this subsection. 

It is illustrated good in the review \cite{Lenz} (see Fig. 5)

\bigskip
 \begin{figure}[h]  
	\label{fig:p28}
	\centering  
	\includegraphics[bb= 140 140  180 180]{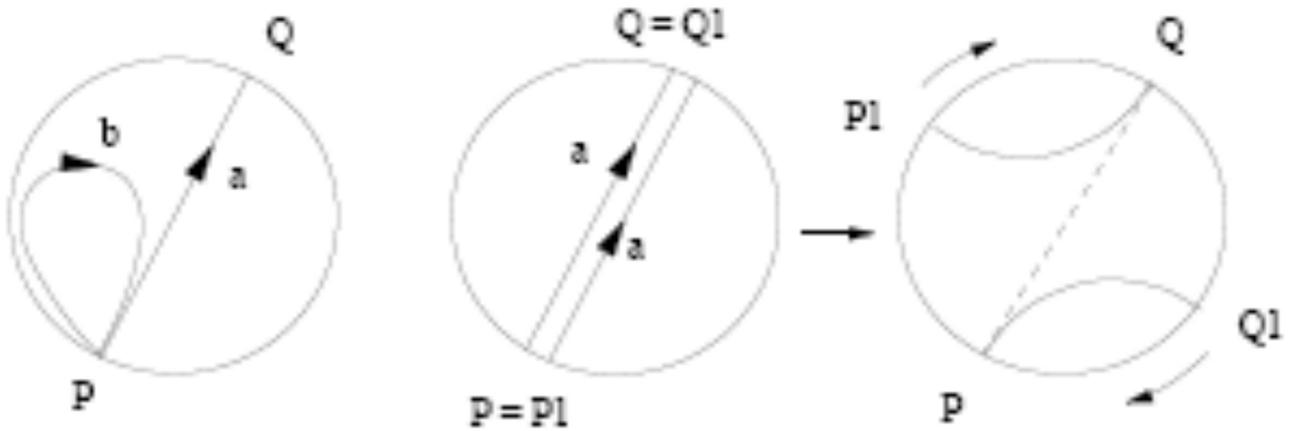}
	\hspace{230.pt}
	\vspace{120.pt}
	\caption{The left figure shows loops $a$, $b$, which on ${\bf RP}^2$ can (b) and cannot (a) be shrunk to a point.
The two figures on the right demonstrate how two loops of the type a can be shrunk to one point. }
	\end{figure}
	
Thus	Fig. 5 demonstates visually the isomorphism 
\be \label{proect}	  \pi_1 SO(3)= \pi_1 {\bf RP}^2={\Bbb Z}_2.\ee
	
	\bigskip It turns out that the Yang-Mills (YM) model, implicating the (initial) $SU(2)$ gauge group, also can be the source of topological defects similar to disclinations \cite{Al.S.} in liquid nematic crystals possessing a one symmetry axis. 	
This was shown in Ref. \cite{Lenz}. Such topological defects are referred to as {\it center vortices} in modern literature. 

If one requires the centre symmetry to be present upon gauge
fixing in the YM model, the isotropical ${\Bbb Z}_2$  group  formed by the centre reflections must survive the "symmetry breakdown" induced
by  eliminating  redundant variables. \par
In
this way, one can change effectively the gauge group:
\begin{equation}
  \label{suso}
  SU(2) \to SU(2)/\mathbb{Z}_2.
\end{equation}
Herewith the centre $\mathbb{Z}_2$ of the gauge group $SU(2)$ consists of two elements \cite{Lenz}: $e$ and $-e$ \footnote{Following \cite{Lenz}, note that the group of centre symmetries divides the set $\cal O$ of $SU(2)$ gauge orbits into two subsets 
${\cal O}_\pm$ corresponding to two eigevalues $c=\pm 1$ of the operator $C_U$ of center reflections. \par
In this case all the gauge fields $A$ can be subdivide into two classes $A_\pm^{ f}$ (if a gauge $f$ is fixed) in such a wise  that the $C_U=1$ transformation maps the subset ${\cal O}_+$ into the subset ${\cal O}_-$. It is equivalent to mapping
$$  Z:~~ A^f_+\leftrightarrow A^f_-        $$
in the space of gauge fields $A$.\par
Herewith since center reflections commute
with the YM Hamiltonian $H$ (due to the above assumption \cite{Lenz} about surviving center reflections), then \cite{Lenz}
$$ [H, Z]=0.  $$
On the quantum level, in this case the operators $H$ and $Z$ possess the common set of eigenfunctions:  
$$  H \vert n_\pm>= E_{n_\pm} \vert n_\pm>, \quad Z \vert n_\pm>=\pm  \vert n_\pm>.       $$ }.

Since 
\be \label{vort}\pi_1\big(SU(2)/\mathbb{Z}_2\big) \simeq \pi_1\big(SO(3))\simeq \pi_1 {\bf RP}^2 =\mathbb{Z}_2,\ee 
the group space of $ SU(2)/\mathbb{Z}_2$ proves to be containing the specific kind of topological defects, referring to as { center vortices}. \par
The isomorphism (\ref{vort}) resembles the (\ref{proect1}) and (\ref{proect}) ones. This points out the similar nature of disclinations \cite{Al.S.} in liquid nematic crystals possessing a one symmetry axis and center vortices \cite{Lenz} in the YM model. 

\medskip
Since ${\bf RP}^2\subset {\bf R}^3$ and 
\be \label{Lenz} \pi_1\big(SU(2)/\mathbb{Z}_2\big)  =\pi_{1} ({\bf RP}^2)= {\Bbb Z}_2,\ee
one can ascertain that the group space  of $ SU(2)/\mathbb{Z}_2$ contains nontrivial singularity lines in ${\bf R}^3$ ("translated" into singularity sheets in ${\bf R}^3$ \cite{Lenz}) similar to those (disclinations) \cite{Al.S.} one discovers in liquid nematic crystals possessing a one symmetry axis  \footnote{Indeed, there is an essential distinction between the cases of liquid nematic crystals possessing a one symmetry axis \cite{Al.S.} and \cite{Lenz} of the YM model possessing the "continuous" $SU(2)$ group geometry simultaneously with the maintained isotropic ${\Bbb Z}_2$ symmetry. 

The sense of this distinction is  that the initial $SO(2)$ (rigid) symmetry inherent in liquid nematic crystals possessing a one symmetry axis is then {\it violated} down to the $O(2)$ symmetry group (that implies the nontrivial degeneration space $R_n= SO(2)/ O(2)\simeq {\bf RP}^2$ \cite{Al.S.}, (\ref{nem}), in the case of such crystals), while the $SU(2)/ {\Bbb Z}_2$ symmetry assumed \cite{Lenz} for the YM model is {\it exact}.

Nevertheless, there is the explicit isomorphism (\ref{Lenz}) between $R_n \simeq {\bf RP}^2$ in the case \cite{Al.S.} of liquid nematic crystals possessing a one symmetry axis and $SU(2)/ {\Bbb Z}_2$ in the case \cite{Lenz} of the YM model possessing the "continuous" $SU(2)$ group geometry simultaneously with the maintained isotropic ${\Bbb Z}_2$ symmetry. }.\par\bigskip 
Transformations associated with such a  singularity, we shall denote them as $U_{\mathbb{Z}_2}(x)$, bear a purely  gauge nature \cite{Lenz}:
$$ A^{\mu}_{\mathbb{Z}_2}(x) = \frac{1}{ig}\,U_{\mathbb{Z}_2}(x)\,\partial ^{\mu}\,U_{\mathbb{Z}_2}^{\dagger}(x). $$ 
The  gauge matrices $ U_{\mathbb{Z}_2}$, written in the cylindrical coordinates $\rho,\varphi,z,t$ as
\be \label{UZ} U_{\mathbb{Z}_2}(\varphi) = \exp{i\, \frac{\varphi}{2}\,\tau^3},\ee
just exhibit the essential properties of singular gauge transformations referring to center vortices and associated with
singular gauge fields. \par
Really, any $U_{\mathbb{Z}_2}$ proves to be singular on the sheet $\rho = 0$ (for all $z,t$) and 
 has the obvious property
 \be \label{discont} U_{\mathbb{Z}_2}(2\pi) = -U_{\mathbb{Z}_2}(0),\ee
i.e. that any such gauge transformation is continuous in $SU(2)/\mathbb{Z}_2$ but discontinuous as an element of $SU(2)$. \par
To make sure that $U_{\mathbb{Z}_2}$ are singular on the sheet $\rho = 0$, it is necessary to consider appropriate {\it Wilson loops} $ W_{{\cal C}, \,\mathbb{Z}_2}$.  \par 
\medskip
Remind herewith (see e.g. \cite{Al.S.}, \S T22) that, in general, Wilson (lines) loops are elements of holonomies groups (isomorphic to the studied gauge groups) with the typical look 
\be
\label{bg}
b_\gamma =P\exp (-\int {\sb \Gamma}  {\bf T} \cdot A _\mu dx^\mu), 
\ee
where the symbol $P$ stands for the parallel transport along the curve $\Gamma$ in the coordinate (for example, the Minkowski) space and $\bf T$ are the matrices of the adjointt representation of the Lie algebra. \par
In Ref. \cite{Lenz} elements $b_\gamma$ were recast to the typical look
\begin{eqnarray}
\Omega\left(x,y,{\cal C}\right)&=& P \exp\left\{-ig \int_{s_0}^{s} d\sigma \frac{dx^{\mu}}{d\sigma}A_{\mu} \Big(x(\sigma)\Big)\right\}= P \exp\left\{-ig \int_{{\cal C}} dx^{\mu}A_{\mu} \right\}.\nonumber\\
\label{PI}
\end{eqnarray} 
Eq. (\ref{PI}) describes a gauge string between the space-time points $x=x(s_0)$ and $y=x(s)$.

$\Omega$ satisfies herewith the differential equation
\begin{equation}
  \label{dipo}
\frac{d\Omega}{ds}= -ig \frac{dx^{\mu}}{ds}A_{\mu} \Omega.
\end{equation}
In this case one can specify $SU(N)$ Wilson loops as \cite{Lenz}
\begin{equation}
   \label{wlop}
W_{\cal C}= \frac{1}{N} \mbox{tr}\, \Omega\left(x,x,{\cal C}\right),    
 \end{equation}
with the trace  taking over the $SU(N)$ gauge group.

For "pure gauges" of the 
\be \label{cl.vac} {\hat A}_i\Rightarrow L^n_i\equiv v^{(n)}({\bf x})\partial_i v^{(n)}({\bf x})^{-1} \quad {\rm as}~ \vert {\bf x}\vert \to\infty; \quad v^{(n)}({\bf x})\in SU(2); \quad n\in {\bf Z},\ee
type the differential equation (\ref{dipo}) can be solved with
\begin{equation}
   \label{wlpg}
\Omega^{\rm pg}\left(x,y,{\cal C}\right) = U(x)\, U^{\dagger}(y) .   
 \end{equation}
In particular, for an arbitrary
path ${\cal C}$ enclosing a center vortex, the appropriate Wilson loop is given as \cite{Lenz} 
\begin{equation}
  \label{wlcv}
W_{{\cal C}, \,\mathbb{Z}_2} = \frac{1}{2}\,\mbox{tr}\, \big\{U_{\mathbb{Z}_2}(2\pi)\, U_{\mathbb{Z}_2}^{\dagger}(0)\big\}= -1.
\end{equation}
The corresponding pure gauge field, got by using the differential equation (\ref{dipo}), has only one non-vanishing space-time component
\begin{equation}
  \label{z2vo}
A^{\varphi}_{\mathbb{Z}_2}(x) = - \frac{1}{2g\rho}\tau^3, \end{equation}
manifestly singular on the sheet $\rho = 0$.
\par
Herewith singular YM fields $A^{\varphi}_{\mathbb{Z}_2}(x)$, given by Eq. (\ref {z2vo}) \cite{Lenz}, represent correctly center vortices in the gauge model involving the $SU(2)/\mathbb{Z}_2$ symmetry group.
\par
\medskip
Knowing singular YM fields $A^{\varphi}_{\mathbb{Z}_2}(x)$, (\ref {z2vo}), the appropriate YM field strength can be calculated with applying the Stokes theorem \cite{Lenz}. 
Then for the flux through an area of an arbitrary size $\Sigma$ located in the $x-y$ plane one gets 
$$\int_{\Sigma}\, F_{12} \rho d\rho d\varphi = -\frac{\pi}{g}\tau^3\, ,$$
and concludes that
$$ F_{12}=  -\frac{\pi}{g}\tau^3 \delta^{(2)}(x).$$
This divergence in the field strength makes these fields irrelevant in the summation over all the configurations. However minor changes, like replacing $1/\rho$ in $ A^{\varphi}_{\mathbb{Z}_2}$, (\ref{z2vo}),
 by a function interpolating between a constant at $\rho=0$ and $1/\rho$ at large $\rho$ eliminate this singularity. The modified gauge field is no longer a pure gauge. Furthermore, a divergence in the action from the infinite field strength can be avoided by forming
closed finite sheets. All these modifications can be carried out without destroying the property (\ref{wlcv}) that the Wilson loop is $-1$ if encloses a vortex.\par
Such "modified" center vortices with the removed ($\delta$-type) singularity at the origin of coordinates were referred to as {\it thick center vorteices} in the review \cite{Jeff1}. Herewith thick center vorteices sweep a surface-like region of
finite thickness and finite field strength.

\medskip Alternatively, {\it Polyakov loops} (lines) can be specified as \cite{Jeff1} Wilson lines winding  once through the lattice in the periodic time direction:
\be
      P({\bf x}) =
    \mbox{Tr}\left[ U_0({\bf x},1)U_0({\bf x},2)...U_0({\bf x},L_t) \right],
\ee 
with $L_t=T^{-1}$ and $ U_0$ being link variables  in the time direction.\par
Maintaining  ${\Bbb Z}_2$ gauge matrices $ U_{\mathbb{Z}_2}$ \cite{Lenz}, (\ref{UZ}), in the YM theory implies that link variables $ U_0$ undergo transformations \cite{Jeff1} 
\be
       U_0({\bf x},t_0) \to z U_0({\bf x},t_0); \quad
 z=\{ \pm 1\} \in {\Bbb Z}_2 \quad
\mbox{for ~all~} {\bf x}.
\label{globalz}
\ee 
Herewith other links (in spatial directions) are assumed to be unchanged \cite{Jeff1}. 

 At these circumstances the transformation law (\ref{globalz}) for link variables $ U_0$ implies the transformation law \cite{Jeff1} 
\be
      P({\bf x}) \to z P({\bf x})
\ee
for Polyakov loops $ P({\bf x})$. \par 
It is easy to understand  \cite{Lenz, Jeff1} that the
centre symmetry (\ref{globalz}) can be realized on the lattice in one
of two ways:
\be\label{im Polyakova}
       \langle P({\bf x}) \rangle = \left\{
       \begin{array}{cl}
         0      &  \mbox{~~unbroken ${\Bbb Z}_2$ symmetry phase;} \cr
\mbox{non-zero} &  \mbox{~~broken ${\Bbb Z}_2$ symmetry phase}.
       \end{array} \right.
\ee

\bigskip  It turns out that  thick center vortices play a crucial role in the confinement of quarks in QCD, as it is understood customary. In Refs. \cite{Lenz,Hugo}  it was argued in favour of this fact.

More exactly, the existence of thick center vortices in the YM  theory involving  the $ SU(2)/\mathbb{Z}_2$ gauge symmetry  being fixed  \cite{Lenz} ensures satisfying  the area law, the main confinement criterion in QCD. 

To ground that the area law is satisfied for Wilson loops $ U_{{\cal C}, {\mathbb{Z}_2}}$, (\ref{wlop}), in that  theory (at $N=2$), it is necessary \cite{Lenz,Hugo} to consider a large area $\cal A$ in a certain plane containing a loop of a much smaller area ${\cal A }_W$.\par 
Herewith the given number $N_1$ of intersection points of (thick) vortices with the area $\cal A$ with those with ${\cal A }_W$ is distributed randomly \cite{Hugo}. \par 
For this random distribution of intersection
points, the probability to find $n$  intersection points in ${\cal A }_W$
 is given by \cite{Lenz, Hugo} 
 $$p_n = \binom{N_1}{n}\Big(\frac{{\cal A}_W}{{\cal A}}\Big)^n\Big(1-\frac{{\cal A}_W}{{\cal A}}\Big)^{N_1-n}\, .$$
 On the other hand, since due to (\ref{wlcv}) \cite{Lenz}, 
 each intersection point contributes a factor  -1, one gets, in the limit of infinite ${\cal A}$ with the density $\nu$ of intersection points kept fixed per area, 
\be \label{arel} \langle W\rangle = \sum_{n=1}^{N} (-1)^n  p_n \to \exp \big( -2\nu {\cal A}_W\big).\ee  
\subsection{What does  it mean, to assign a spin structure to a manifold?}
Considering a flat (in particular, Minkowski) space, one would take in his mind that it is the ordinary Hausdorffian, paracompact and connected space  (say, $\pi_0~M=0$ for the Minkowski space $M$) of the class $C^{\infty}$.

Let us consider now the space $\cal F$ each point of which represents an isotropic flag in the fixed point of the general bend space-time  manifold $\cal M$. Such a  space $\cal F$ is called \cite{Penrous1} {\it the  beam of isotropic flags}  of the space $\cal M$ (see Fig. 6). It is, indeed, a 8-dimensional  space since alone the space $\cal M$ is four-dimensional while the space ${\cal F}_P$ of sotropic flags in an arbitrary point $P$ of the manifold $\cal M$ is also four-dimensional (since  appropriate isotropic vectors $K$ are four-dimensional, as it was discussed above).  Isotropic flags in the point $P$ can be understood as objects in the {\it tangential space} in the point $P$. The latter one  is the vector Minkowski space. 1
\begin{figure}[h]  
	\label{fig:p29}
	\centering
	 \vspace{150.pt} 
	\includegraphics[bb= 130 130  160 160]{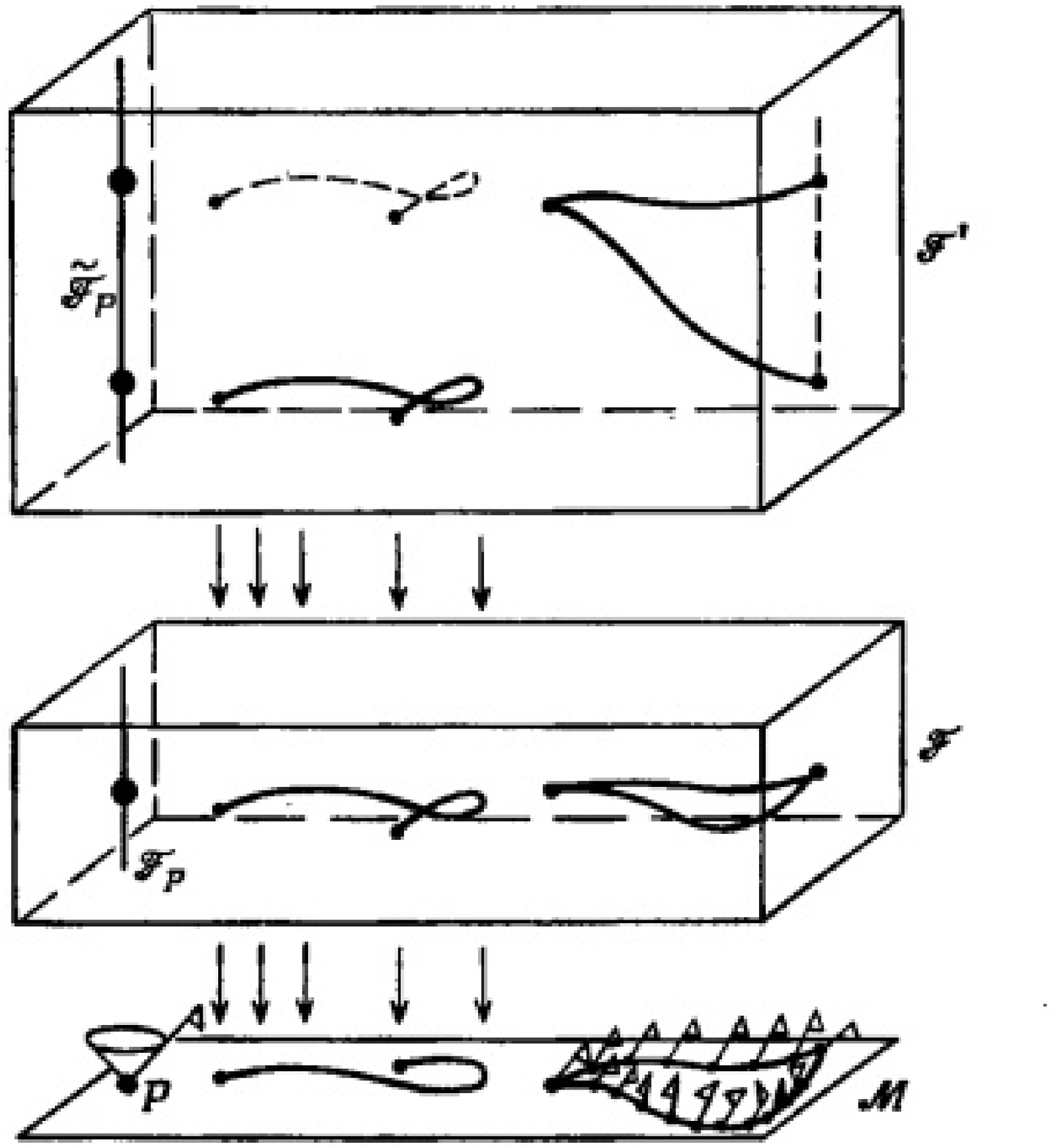}
	\hspace{240pt}
	\vspace{150pt}
	\caption{ The beam of isotropic flags $\cal F$ on $\cal M$ and its two-fold covering space, the beam of spin-vectors ${\cal F}'$.}
	\end{figure}
\newpage
Thus for existing the space $\cal F$, two global restrictions on  $\cal M$ are necessary. Firstly, isotropic flags are connected with only a one of two light half-cones in the { tangential space} in the point $P$: namely with that one directed {\it in the future}.  Therefore it is  necessary to have the possibility to choose the light half-cones in concord in the whole $\cal M$.

In other words \cite{Penrous1}, 
 {\it the manifold}  $\cal M$  {\it should be oriented in the time}. 
 
 Secondly, the choice of  the space-time orientation is required   for the algebra of spin-vectors in  each point. It is so  since the multiplication on $e^{i\theta}$ should implicate a rotation of  isotropic flags in the definite direction. The fact that this requires the namely the definite {\it space-time} orientation and not a definite {\it spatial} orientation follows from the circumstance that a positive rotation of isotropic flags  allots the sphere $S^+$ with a  positive orientation; respectively,  the sphere $S^-$ is allotted with a  negative orientation.
 
Therefore it is  necessary to have the possibility to  choose, in concord and continuously, the space-time orientation in the whole $\cal M$. Thus \cite{Penrous1} {\it 
the manifold}  $\cal M$  {\it should be oriented in the space and time}. 

\medskip But if we desire to go over from the notion   ``isotropic flag'' to  the notion ``spin-vector'', two  above restrictions are insufficiently.  The manifold  $\cal M$ should also permit the possibility to specify therein a {\it spin structure} \footnote{The question about the existence of  the spin structure on the manifold  $\cal M$  differ from the question about the existence of  some (for instance, nonzero) spinor fields on $\cal M$. The latter one is similar to the question either a nonzero vector field exists on a two-sphere. But in  the absence of a spin structure the alone notion of a global spinor field becomes meaningless. }, i.e., roughly speaking, an instruction allowing to  trace the sign of the  spin-vrctor  not only in the case it is rotated in the fixed point of the manifold  $\cal M$, but also when it is moved from a point to a point  within $\cal M$. 

If the manifold  $\cal M$ is topologically simple, the pointed spin structure exists and is unique. But if $\cal M$ is topologically nontrivial, it can both permit and not permit a co-ordinated spin structure; herewith  in the case when such spin structure exists, it can be  unique or cannot be  unique.  In a general case, it turns out  that the conditions ensuring the existence and   uniqueness of the  spin structure depend on its  topology and does not depend on the look of its (Lorenz) metric. 

\medskip According to the  said above, we claim now that the space $\cal F$  possesses the appropriate two-fold covering space ${\cal F}'$, Fig. 6, which will be actually the  space of spin-vectors on $\cal M$ \footnote{In contrast to a universal  covering space, a general covering space should satisfy only the claim of connection, and it should be mapped into the initial space in such a wise that the local topology is maintained and the inverse  map of the point is a discrete sequence of points (the latter claim enters the standard definition for a covering: see e.g. Lecture 2 in \cite{Postn4}).}. 

The space $\cal F$ should be ``appropriate'' in the sense of its reducing to $\tilde {\cal F}_P$, the universal  covering space of   the space $ {\cal F}_P$ over an arbitrary point $P\in {\cal M}$.  

It can be assumed that the universal  covering  space $\tilde {\cal F}$ for $ {\cal F}_P$ satisfies this condition in  the natural way (i.e. ${\cal F}'=\tilde {\cal F}$), but since the complete ``turn'' of   the space ${\cal F}$  includes also the  turn of   the space ${\cal M}$, this condition can also to be not  realizeed.  

And moreover, the situation proves to be more complicate. We shall see now that in fact two somewhat another  obstacles to existing  ${\cal F}'$ are possible. The first of them is connected with the question either the space ${\cal M}$ is one-connected or not, while the second one arises only in the case of a multi-connected ${\cal M}$. 

Really, let us consider closed contours in ${\cal F}$ and their projections on ${\cal M}$. The projection from ${\cal F}$  on ${\cal M}$ maps a flag in  the point $P$ on this point $P$; thus any space ${\cal F}_P$ maps entirely in  the unique point $P$ (see Fig. 6). An arbitrary way in ${\cal F}$ is projected into a way in  ${\cal M}$; it is obvious that a closed contour in ${\cal F}$ is projected herewith into a  closed contour in  ${\cal M}$. Any way in ${\cal F}$ corresponds to such a motion which moves any isotropic  flag in ${\cal M}$ and which returns finally this flag (in the case of a {\it closed} way) in its initial position. The projection describes merely the motion of the base point in ${\cal M}$.

A contour in ${\cal F}$ lying entirely in the universal covering $ {\cal F}_P$ at a fixed $P$ is projected  into a topologically trivial contour (the point $P$) in ${\cal M}$ \cite{Postn4}. As we have seen in  the previous section \cite{Penrous1}, there exist two classes (I and II) of  closed contours in ${\cal M}$. 

The of the {\it first } type obstacle arises in  the case of a nontrivial topology in the manifold ${\cal M}$ is associated with the fact that pointed two classes of contours can merge into a one fixed type, namely into an of  the  class I contour in an arbitrary $ {\cal F}_P$, which  cannot be contracted into a point. Herewith upon the deformation process within ${\cal F}$, such a contour can return into $ {\cal F}_P$ as that belonging to the class II of contours can be contracted into a point. 

In this case no spin-vector on ${\cal M}$ can exist. Really, let us assume  that such spin-vectors exist, and  let us consider a contour $\lambda$ of  the class I on  the fixed space $ {\cal F}_P$, is set merely by the rotation of the flag cloth, for a given isotropic flag, onto the angle $2\pi$. As a result, the appropriate  spin-vector ${\bf \kappa}$ is mapped into $-\kappa$.  Any closed contour in ${\cal F}$ in which the contour $\lambda$ can be transformed transfers continuously a nonzero  spin-vector into its opposite. But if $\lambda$ can be turned continuously into a sole point  in ${\cal F}$, then the appropriate spin-vector should be equal to its opposite (in other words, $\kappa=-\kappa=0$, see Subsection 2.2).  Therefore it is impossible to introduce a nonzero  spin-vector in the manifold ${\cal M}$. 

\medskip Let us assume now that the first type obstacle is absent. Then in the case of the manifold ${\cal M}$ containing a   contour $\gamma$ cannot be contracted into a point, i.e.  in the case $\pi_1 {\cal M}\neq 0$, the {\it second} type obstacle can arise. If an isotropic flag moved along the contour $\gamma$ return in its initial position $P$, then the appropriate spin-vector $\kappa$ should return either in its initial value or to $-\kappa$.  Thus one should choose the one of these two possibilities. 

If the contour $\gamma$ is such that a multiple contour $m\gamma$ (i.e. $\gamma$ passing $m$ times) is not contracted into a point, any of the mentioned two possibilities can be realized equally, but  they lead to different spin structures on ${\cal M}$ (it is assumed that the spin structure is not excluded by another contours). 

In this case two alternatives will enter the definition of a spin-vector. By the choice has bein made for a $\gamma$,  the  choice is specified for all the contours in ${\cal F}$ can be either projected on $\gamma$ or deformed in $\gamma$ on ${\cal M}$. 

Let us assume now that the contour $\gamma$ is such that an odd  multiple contour $m\gamma$ can be contracted into a point on ${\cal M}$. Then the contour $m\lambda$ is deformed in a contour on a  $ {\cal F}_P$ for a contour $\lambda$ in $ {\cal F}$ if  the contour $m\gamma$ is deformed in the point $P\in {\cal M}$.  

If the latter contour on $ {\cal F}_P$ belongs to the class I, then the spin-vector $\kappa$ taken near $m\lambda$ should be transferred continuously into $-\kappa$; if it  belongs to the class II, the spin-vector $\kappa$ should be transferred into $\kappa$.  Since $m$ is odd, $\lambda$ is fixed by this condition as that transferring $\kappa$ into $-\kappa$ or $\kappa$, respectively, and herewith  unambiguous. 

\medskip Finally, it can turn out that while all the odd contours $\gamma$ cannot be  contracted into a point, some (minimum) {\it even} multiple contours $2n\gamma$ {\it can be}  contracted into a point. Then it must be one of two alternatives. Either  all the corresponding contours $2n\lambda$ on $ {\cal F}$ are transferred into the  contours of the class II on $ {\cal F}_P$ at the defformation of the contour $2n\gamma$ in  the point $P$ or some of them (and then all they, actually) are transferred into the  contours of the class I. 

In the first case the spin-vector $\kappa$ moving continuously along $2n\gamma$ should be  transferred into itself. Therefore {\it each} of two possibilities $\kappa\to\pm \kappa$ is acceptable for the  single passage   of the contour $\lambda$, and we come  (as it was earlier)  to two possible spin structures on ${\cal M}$  (if, of course, the spin structure is not excluded by another contours). 

 Let us assume however that the contour $2n\gamma$ is transferred into a  contour of the class I on $ {\cal F}_P$: hence follows the claim  $\kappa\to- \kappa$ along the contour $2n\gamma$. Then {\it any} of  two  possibilities $\kappa\to\pm \kappa$ is not acceptable near $\lambda$, and in  this is the essence of the second type obstacle for ${\cal M}$ to permit the spin structure.
 
  In contrast to the first type obstacle, the second type obstacle can arise only when $\pi_1 {\cal M}\neq 0$, and it disappears (in contrast to the first type obstacle) at going over to the universal covering of ${\cal M}$. 
  
 \medskip It is possible to construct the patterns of space-time models \cite{mod} in which a one of abovementioned obstacles arises and which satisfy nevertheless the conditions to be oriented  in the  time and in the space-time. Herewith such  models do not seem to be senseless from the physical point of view. Actually, one deals here with the display of more general thesis correct for manifolds of an arbitrary dimension.
  
There exists a topological invariant called \cite{Penrous1} {\it the second class $\omega_2$ by  Shtiffel-Whitney},  equality of which to zero in the case of an oriented manifold ${\cal M}$ is the necessary and sufficient condition for the statement \cite{Penrous1} {\it that the manifold ${\cal M}$ possesses the spin structure}. More exactly, it is the necessary and sufficient condition for the existence of the common (however   two-digit specified) spinor objects on ${\cal M}$ \cite{Mil}. 

The  condition $\omega_2=0$ can be formulated strictly as following \cite{Penrous1}.

{\it On an arbitrary closed two-surface ${\cal G}\in {\cal M}$ $({\rm dim}{\cal M}\geq 3)$, there exists the system of $n-1$ continuous fields of  tangential  vectors to ${\cal M}$ linearly  independent in each point of  ${\cal G}$. If the manifold ${\cal M}$ is oriented (that is expressed by the condition $\omega_2=0$), one can replace the number $n-1$ with  the number $n$}. 
 
 One can demonstrate \cite{Penrous1} that if this condition (call it Ccondition A) is fulfilled for a  space-time manifold ${\cal M}$ (oriented  in the  time and in the space-time; then $n=4$),   the above obstacles are absent. 
 
 Let us consider  preliminary the rotation group $SO(4)$ and let us show that, like the $SO(3)$ case, the  closed ways in  $SO(4)$ are split into the classes I and II  such that a double way of  the class I  is the way of  the class II: in other words, that $\pi_1 (SO(4))={\Bbb Z}_2$ \footnote{The same is correctly for $SO(n)$ ($n\geq 3$).}.  
 
 We utilize the quaternion theory (see e.g. \S 2 to Chapter 1 in \cite{Penrous1} or \S A8 in \cite{Al.S.}) and note that any element of $SO(4)$ can be got by means of acting onto the unit quaternion $\bf q$:
 \be \label{quaternion} 
 {\bf q}\longmapsto \tilde{\bf q}={\bf aqb},
 \ee
with $\bf a$ and $\bf b$ being  the fixed  unit quaternions. This follows from the fact \cite{Al.S.} that $\tilde{\bf q}\tilde{\bf q}^*={\bf q}{\bf q}^*$ is the four-dimensional Euclidian norm, while the complete  dimension of $SO(4)$, ${\rm dim}~SO(4)=6$, is got at the action (\ref{quaternion}). 

There is the ambiguity 
$$  ({\bf a},{\bf b})=(-{\bf a},-{\bf b}), $$
but at disregarding this, the pair $({\bf a},{\bf b})$ is determined in the unique way by  the element of $SO(4)$ representing by this pair. 

\medskip Let us assume also that the four-dimensional  manifold ${\cal M}$ is oriented  in the  time and in the space-time and that Condition A is satisfied for ${\cal M}$.  Let us imagine that the tangential space $T_p$ in  each point $P\in {\cal G}$  is mapped linearly on ${\bf R}^4$ in such a wise that four linearly  independent vectors in $P$ are  mapped, respectively, in  four coordinate vector bases in ${\bf R}^4$. In other words, we consider  four vector fields figured in Condition A as  coordinate  axes in each point of ${\cal G}$. 

The light future cone in a point $P$ will be mapped into the half-cone $K^+$ in ${\bf R}^4$ (see Fig. 7).  The one of the main axes of the half-cone $K^+$ is the map $\bf A$ in ${\bf R}^4$ of a future-directed time-like vector in $T_P$ (more precisely, it is the axis lying {\it inside} $K^+$; see Fig. 7). At a transition of the point $P$ along ${\cal G}$, the vector ${\bf A}\in {\bf R}^4$ moves  continuously with $P$.

\begin{figure}[h]  
	\label{fig:p30}
\vspace{140.pt}	
	\centering  
	\includegraphics[bb= 140 140  180 180]{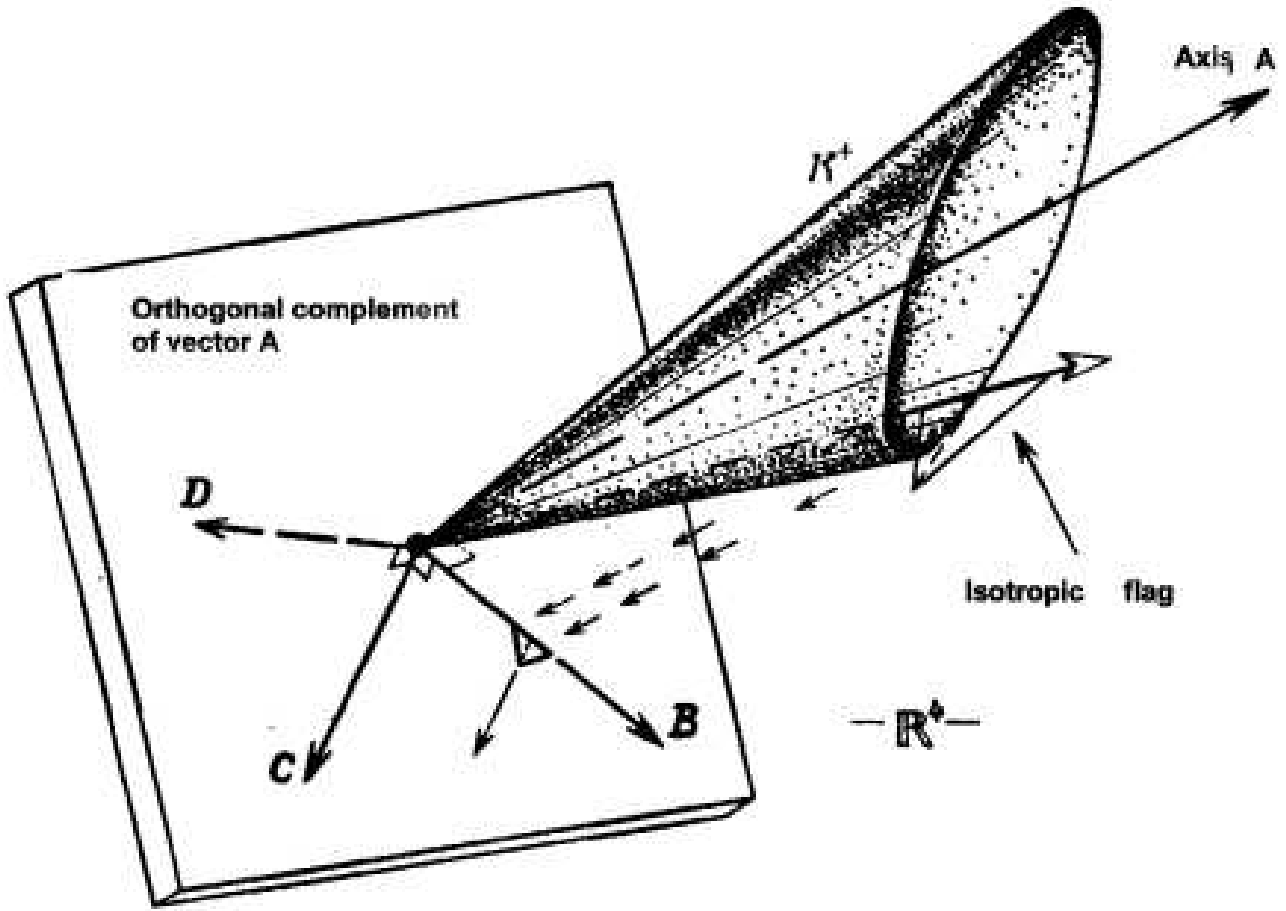}
	\hspace{240pt}
	\vspace{200pt}
	\caption{ A continuous map  onto  ${\bf R}^4$ of the isotropic future cone and isotropic flag results in the unique wise the definite coordinate system ${\bf ABCD}$ which is the right and orthonormalized with respect to the Euclidian metric in ${\bf R}^4$.}
	\end{figure}

Let us consider now the isotropic flag  in the point $P$. Its image in  ${\bf R}^4$ will be the ``flag'' whose  flagstaff  indicates the direction of the half-cone $K^+$ director and whose cloth is tangential to $K^+$. 

Let $\bf B$ be the projection of this flagstaff orthogonal to the  axis $\bf A$ (with respect to the Euclidian metric in ${\bf R}^4$). The projection of the {\it flag cloth} orthogonal to the  axis $\bf A$ contains only one direction $\bf C$ perpendicular to the vector ${\bf B}$ (and ${\bf A}$). Let us choose  the vector ${\bf D}$ in such a wise that it supplements the vectors ${\bf A}$, ${\bf B}$, ${\bf C}$  to the right tetrad, and let us  normalize all the vectors ${\bf A}$, ${\bf B}$, ${\bf C}$ and ${\bf D}$ in such a wise that they become unit vectors (in the Euclidian metric in ${\bf R}^4$). 

Thus we, in the  continuous wise, ascribe the orthonormalized right coordinate system ${\bf ABCD}$ to each  isotropic flag on an arbitrary point of ${\cal G}$ (i.e. to each point of the space $\cal F$ over ${\cal G}$). 

Note  that the achieved correspondence possessess the following property: if an  isotropic flag describes a way of the class I (or II) with the fixed point $P$,  then the appropriate coordinate system ${\bf ABCD}$ performs the continuous rotation in $SO(4)$ of the class I (or II).  To show  this, it is necessary to consider the rotation of the flag cloth onto the angle $2\pi$ and then conclude according to the  continuity reasoning. 

\medskip Let us analyse now two types of possible obstacles  to the existence of spinors in the space-time manifold ${\cal M}$ oriented in the space and time. In the case of  classes I and II flowing together when the contour $\lambda$ in ${\cal F}_P$ corresponding to the rotation  onto the angle $2\pi$ is deformed in a point on ${\cal F}$, its projection on ${\cal M}$ forms the closed surface ${\cal G}$, to which Condition A is applicable.  If the abovementioned systems of  vectors exist on ${\cal G}$, one can describe, in a continuous wise, the orientation of the investigated flag by utilizing the system ${\bf ABCD}$ in ${\bf R}^4$ as it was pointed above. Any position of a contour in ${\cal F}$ corresponds then to a continuous motion of the system ${\bf ABCD}$ in ${\bf R}^4$; in particular, the initial one, to the  rotation  on $2\pi$, while the final one (continuous with the initial one), to the absence of any rotation, that is impossible.  Thus if Condition A is satisfied on ${\cal M}$, the contour $\lambda$ in ${\cal F}$ cannot be contracted into a point in ${\cal F}$;  this means that the of the first  type  obstacle cannot arise. 

\medskip With theaid of similar reasoning, the second possibility for the absence of a spin structure is excluded. The disagreement  in the transport of the flag upon $2n$ revolutions along the contour $\gamma$, i.e. two  revolutions along the contour $\eta=n\gamma$. Due to assuming,   the contour $2n\gamma=2\eta$ should be contracted into the point $P\in {\cal M}$. Course  such contracting, this contour forms the closed surface in  ${\cal M}$ ``welded on'' to the contour $\eta$.  Condition A can be applied to such surface, and the  coordinate system ${\bf ABCD}$ in ${\bf R}^4$ can be utilized for mapping the flags transferred along the contour $2\eta $ in the different stages of deforming it in a point. 

The considered obstacle  arises now in that case when the contour $2\zeta$ on ${\cal F}$, where  $\zeta$ is projected into $\eta$, is deformed in the contour of the class I in ${\cal F}_P$. However a flag transferred along the contour $2\zeta$ in ${\cal F}$ can be represented by the {\it double} motion of the system ${\bf ABCD}$ and therefore  by the contour of the class II in $SO(4)$. If a finite  contour belongs to the class I, the appropriate initial contour of the class II in $SO(4)$ would be deformed continuously in to the way of the class I therein, but  this is impossible; thus  this  obstacle  also cannot arise. 

\medskip In the case when \cite{Penrous1} {\it all three} properties (to be oriented in the time, in the space-time and to have a spin structure), one utilizes the  term {spinoral structure} instead of the conventional term spin structure.  Thus if ${\cal M}$ possessess a spinoral structure, there exists a spinoral system on ${\cal M}$ (based onto the isotropic flags and spin vectors). In other words,  there exists the space ${\cal F}'$, specified above (it covers twice the space ${\cal F}$ of isotropic flags),  If the manifold  ${\cal M}$ is one-connected ($\pi_1 {\cal M}=0$), ${\cal F}'$ will be, in fact,  the universal covering $\tilde F$ \footnote{In each space $\tilde{\cal F}_P$, the way between two points  representing the unique point in ${\cal F}_P$ corresponds to the rotation onto the angle $2\pi$; this ensures the fulfillment of the same property for ${\tilde F}$.}. 

\subsection{Fermions with spin $1/2$ as global $SO(3)$ vortices.}
Introducing the spin (spinoral) structure in the Minkowski  space $M$ (that is equivalent to introducing therein isotropic flags and appropriate spin vectors \cite{Penrous1}) allows to describe correctly fermions possessing the spin $1/2$. 

This model functions good and effectively when massless (anti)neutrinos are in the question, but also in the case of massive electrons the arguments \cite{Penrous1} remain valid  in the part choosing the pair $(\xi,\eta)$ as the components of the flag cloth, while its   flagstaf is now a time-like 4-vector.  Herewith the {\it complex} pair $(\xi,\eta)$ permits the trasperent interpretation in theoretical physics as the (two)-components of a bispinor (in the terminology \cite{Penrous1}, it is a spin-vector $\kappa$), while the model \cite{Penrous1} ensures good relativistic properties of bispinors. 

On the other hand, Eq. (\ref{minus}) permits the treatment of the fields $\xi$  and $\eta$ (undertaken individually) as those possessing the spin $1/4$.  There are not physical fields (for instance, in QED these fields are manifestly $U(1)$ covariant; besides that, the pair $(\xi,\eta)$ is always covariant with respect to spin transformations (\ref{spt})). 

\medskip The next important lesson we learn from the present study, based essentially on the  flag model   \cite{Penrous1} (specifying correctly  the spin [spinoral] structure on the Minkowski space $M$) and on  that topological specific of the global $SO(3)$ group that its group space is two-connected, $\pi _1 SO(3)={\Bbb Z}_2$. The latter fact, in  turn, involves the  presence of two classes, I and II, of contours  in   the $SO(3)$ group  space.  Herewith while contours of the class I are topologically nontrivial and corrrespond to the degree of map 1, all the contours of the class II ate topologically trivial and can be contracted into a point (they corrrespond to the degree of map 0). 

In the present study examples of ${\Bbb Z}_2$ vortices were considered. There are disclinations  \cite{Al.S.} in nematic crystals possessing  a one symmetry axis directed along the axis $z$ of the studied crystal and center vortices \cite{Lenz} in the Yang-Mills ($SU(2)$) model.

These examples of ${\Bbb Z}_2$ vortices suggest the idea that contours of the class I \cite{Penrous1} in the  $SO(3)$ group space also can be treated as (global) $SO(3)$ vortices. 

\medskip In Ref. \cite{A.I.} there was given an enough transparent interpretation of the property of a two-component spinor 
 \be\label{varphi} \varphi= \left (
\begin{array}{llcl} 
 \varphi_1\\ \varphi_2	
\end{array}\right )\ee
that at the rotation onto the angle $2\pi$, $\varphi(2\pi)=-\varphi$.

The proof of this fact is easy (see \S 7 in \cite{A.I.}). 

At an infinitesimal  rotation of the (three-dimensional) coordinate system onto the angle $\delta \theta$, the spinor $\varphi$ undergoes the infinitesimal $SO(2)$ (global) transformation 
\be \label{fio2}
\varphi\to \varphi'= \varphi+\delta\varphi; \quad \delta\varphi=-i{\bf s}\delta \theta\varphi,
 \ee
with ${\bf s}=(\hbar/2) {\bf \sigma}$ being the  spin operator ($\bf \sigma$ are the Pauli matrices).
 
 At a rotation onto a finite angle $ \theta$ around the axis whose direction is specified with the unit vector $\bf n$,
\be \label{fioc2} 
\varphi'=e^{-i\frac \theta 2 {\bf n}{\bf \sigma}}\varphi.
\ee 
Since 
\be \label{Pauli} ({\bf n}{\bf \sigma})^{2k}=1; \quad  ({\bf n}{\bf \sigma})^{2k+1}={\bf n}{\bf \sigma} \quad (k\in {\bf Z}),\ee
Eq. (\ref {fioc2}) can be represented as
\be \label{fioc22} 
\varphi'= (\cos\frac \theta 2 -i {\bf n}{\bf \sigma}\sin \frac \theta 2)\varphi.
\ee
To derive this Eq., it is enough to expand in the power series $\cos(\frac \theta 2{\bf n}{\bf \sigma})$ and $\sin(\frac \theta 2{\bf n}{\bf \sigma})$ with account of the  relations (\ref{Pauli}). 

Just from (\ref{fioc22}) it follows \cite{A.I.} that 
\be \label{pi2} 
\varphi'(2\pi)=-\varphi(0).
\ee

\medskip Deriving \cite{A.I.} Eq. (\ref{pi2}) in the nonrelativistic three-dimensional case, implicating $SO(3)$ (global) rotations, can be generalized  easy to the relativistic four-dimensional case. 
Then instead of the  spinor compnents $\varphi_1$ and $\varphi_2$, complex numbers $(\xi,\eta)$, a la \cite{Penrous1}, ``enter the game'' in such a wise  that Eq. (\ref{minus}) is fulfilled, generalizing Eq. (\ref{pi2}) \cite{A.I.} \footnote{It can be argued (see e.g. \S 21 in \cite{BLP}), that a bispinor $u$, consisting of four spinor components, turns into a spinor $\varphi$, (\ref{varphi}), at small velocities  $v\ll c$. In this case one can neglect the momentum of a fermion (with the spin 1/2) in the Dirac equation. Then its energy is $E\to mc^2$, and as the consequence, two components of a  bispinor coincide  with each other.  }. 
\begin{appendix}
\renewcommand{\theequation}{A.\arabic{equation}}
\setcounter{equation}{0}
\section{Appendix 1. Why $O_+^\uparrow (1,3)$ is a six-dimensional space?}
$O_+^\uparrow (1,3)$ is the denotation \cite{Penrous1} for the {\it restricted Lorenz group}, including the matrices with ${\rm det}~O_+^\uparrow (1,3)=1$, while $(O_+^\uparrow (1,3))_0^0\geq 1$.  

The unimodular matrices $\bf A$,  (\ref{unimod}), satisfy this criterion as it was shown in  \cite{Penrous1}. It is most simply to prove the  assertion that any restricted Lorenz transformation corresponds to only two spin unimodular transformations (\ref{unimod}) the one of which is opposite to another. 

This follows from the fact that the Lorenz group,  figuring actually in Eq. (\ref{spin transformation}), should have the dimension six. Really, the  spin-matrices (\ref{unimod}) form indeed the six-dimensional (i.e. the complex three-dimensional) system $A_x, A_y, A_z$. And moreover,  only the discrete set of spin-matrices (concretely, only two spin-matrices) corresponds to a one Lorenz transformation.  This complete subgroup should contain the entirely connnected component of the Lorenz group including  the identical transformation. 

Alternatively, this assertion can be proved by means of explicit constructing spin-matrices corresponding to a ground Lorenz transformation from which the whole Lorenz group can be formed. Spatial rotations and Lorenz busts belong to such ``ground  transformations''.  As it is well known, Lorenz busts can be represented as \cite{Penrous1} 
$$ \tilde T= (1-v^2)^{-1/2} (T+vZ); \quad  \tilde X=X; \quad  \tilde Y=Y; \quad \tilde Z= (1-v^2)^{-1/2} (Z+vT),$$
with $v$ being the velocity parameter. 

Any restricted Lorenz transformation can be formed from an (eigen) spatial rotation,  a  Lorenz bust in  the $z$ direction and the second spatial rotation. 

Let us elucudate how such a transformation is characterized by its action onto a Minkowskian tetrad. We choose the first spatial rotation in such a wise that it transfers the vector ${\bf z}$ in the space-time plane containing both the initial and final directions of $t$. Then the bust imparts to the vector ${\bf t}$ its final direction, while the second spatial rotation is utilized for the proper orientation of  the vectors ${\bf x},{\bf y}$ and ${\bf z}$. 

Thus it remains to show only that spatial rotations and $z$-busts can be got from spin transformations. 

\medskip Let us consider, to begin with, rotation and let us establish the following result \cite{Penrous1}. 

{\it Any unitary spin  transformation corresponds to the unique eigen rotation of  the sphere $S^+$; inversely, any eigen rotation of  the sphere $S^+$ corresponds to the  only two unitary spin  transformations any of which is opposite to another.   }

First of all, let us consider what a geometrical sense of our transformations. The Lorenz transformations can be treated as active in this case. The spheres $S^+$ and $S^-$ are considered herewith as the part of  the coordinate system and {\it do not participate} in the transformation;  thus at a shift of any isotropic future (past) direction, its representation on $S^+$  ($S^-$) is also shifted. For example, a $({\bf x},{\bf y},{\bf z})$ rotation remaining $t$ invariant, corresponds to the rotation of  the image on $S^+$  ($S^-$). 

The plane $\Sigma$ (Fig. 2) is also the part of  the coordinate structure, and it remains invariant, while images on this  plane of  isotropic straight lines $\zeta$  are shifted (for instance, because of (\ref{xyz1}), (\ref{bfK})).  In the latter case one speaks about ``motions'' of  the plane $\Sigma$ \footnote{Of course, $S^+$,  $S^-$ and $\Sigma$ are invariant not to the larger degree than different coordinate hyperplanes: vectors  lying  in these hyperplanes exceed their  boundaries	in a general case upon performing Lorenz transformations.}. 

It is important to remember that although one deals here with the representation of solely isotropic  directions of the space $M$ and that the transformations of {\it all the} vectors in $M$ are determined by the transformations of  these isotropic directions.

\medskip It follows  from (\ref{spin transformation})  that the variable $T$ is invariant with respect to the unitary spin  transformation since its trace ($=2T$) is always invariant at unitary   transformations (with the equal success, one can refer to the invariance of the expression $\xi\bar\xi+ \eta\bar\eta$, representing correctly the Hermitian norm of  the pair $(\xi,\eta)$). 

The restricted Lorenz transformations at which the variable $T$ is invariant are merely eigen rotations of  the sphere $S^+$ (since  they maintain $X^2+Y^2+Z^2$).

\medskip To demonstrate explicitly the  inverse assertion, note firstly that any eigen rotation \linebreak $({\bf x},{\bf y},{\bf z})\to ({\bf x}',{\bf y}',{\bf z}')$ of  the sphere $S^+$ can be comprised of sequential rotations around the axes $Y$ and $Z$. Really, the  tetrad $({\bf x}',{\bf y}',{\bf z}')$ is specified by the polar coordinates $\theta,\phi$ of  the axis ${\bf z}'$ regarding $({\bf x},{\bf y},{\bf z})$ and the angle $\psi$ formed by the planes $({\bf x}',({\bf z}')$ and $({\bf z},({\bf z}')$ (the mentioned three angles are well-known Euler angles).   Thus the quested transformation will be achieved by the rotation onto  the angle $\psi$ around the vector $\bf z$,   then by the rotation onto  the angle $\theta$  around the initial vector $\bf y$ and, finally, by the rotation onto  the angle $\phi$ around the initial vector $\bf z$. 

\medskip Let us demonstrate now how these elementary rotations can be represented by unitary spin transformations. It will follow hence  that any eigen rotation of  the sphere $S^+$ can be represented by an unitary spin transformation since the product of unitary matrices is always an unitary matrix. 

It is obwious that the  rotation of  the sphere $S^+$ around the axis $z$ onto  the angle $\psi$ arises from the rotation of  the Argand plane relatively to the origin of coordinates onto  the angle $\psi$. Such a rotation is set by the relation 
\be \label{arg}
\tilde \zeta =e^{i\psi} \zeta,
 \ee
 i.e. by the spin rotations
\be \label{spr} \left (
\begin{array}{llcl}
\tilde \xi\\ 	\tilde \eta
\end{array}
\right ) =\pm \left (
\begin{array}{llcl} e^{i\psi/2} \quad 0\\ 0 \quad e^{-i\psi/2}
\end{array} \right )\left (\begin{array}{llcl}
 \xi\\ 	 \eta
\end{array}
\right ).
\ee 
Then we assert that the  rotation of  the sphere $S^+$ around the axis $y$ onto  the angle $\theta$ is set by the following unitary spin transformations: 
 \be \label{spr1} 
 \left (
\begin{array}{llcl}
\tilde \xi\\ 	\tilde \eta
\end{array}
\right )=\pm \left (\begin{array}{llcl} \cos\theta/2\quad -\sin\theta/2 \\\sin\theta/2\quad\cos\theta/2\end{array}
\right )\left (\begin{array}{llcl}
 \xi\\ 	 \eta
\end{array}
\right).
 \ee
 Since the transformations (\ref{spr1}) are unitary,  they represent  undoubted {\it a rotation}. Moreover, since the difference $\xi\bar 	 \eta-\eta\bar \xi$, as well as the sum $\xi\bar 	 \xi+\eta\bar \eta$, are invariant, it follows from (\ref{xyz1}) that $y$-coordinates of points on $S^+$ are invariant under  (\ref{spr1}).  Therefore, the considered rotation proceeds around the axis $y$. 
 
 Finally, any  transformation (\ref{spr1}) transfers the point $(1,0,0,1)$ into  the point $(1,\sin\theta,0,\cos\theta)$;  thus the rotation angle is really equal to $\theta$.
 Similarly, it is possible to show that the unitary spin transformation 
  \be \label{sprx} 
 \left (
\begin{array}{llcl}
\tilde \xi\\ 	\tilde \eta
\end{array}
\right ) =\pm  \left (
\begin{array}{llcl} \cos \xi/2 \quad i\sin\xi/2\\ i\sin\xi/2 \quad \cos \xi/2\end{array}
\right )\left (\begin{array}{llcl}
 \xi\\ 	 \eta
\end{array}
\right)
 \ee 
 corresponds to the rotation around the axis $x$ onto  the angle $\xi$.
 
 Thus our asumption \cite{Penrous1} is  proven completely. 
\end{appendix}

\end{document}